\documentclass[11pt,a4paper]{article} 
\pdfoutput=1
\usepackage{jheppub, hyperref}
\usepackage{graphicx,xcolor}
\usepackage{color}
\usepackage{slashed,phonetic}
\usepackage{hyperref}
\usepackage{multirow}
\usepackage{appendix}

\usepackage{stmaryrd,fdsymbol} %https://it.overleaf.com/project/60226a0903a710037463e4d2
\newcommand\Indalo{$\mathrel{\ooalign{\raise-.3ex\hbox{$\Yup$}\cr
  \hidewidth\raise.5ex\hbox{$\upblackspoon$}\cr
  \hidewidth\raise.5ex\hbox{$-\mkern-1.5mu$}\cr
  \hidewidth\raise1.1ex\hbox{$\cap\mkern-1.4mu$}\cr}}$}
\title{Minimal SU(5) Asymptotic Grand Unification}

\author[1]{Giacomo Cacciapaglia,}
\author[2]{Alan S. Cornell,} 
\author[1]{Corentin Cot,} 
\author[1]{Aldo Deandrea} 
\affiliation[1]{Universit\'e de Lyon, F-69622 Lyon, France: Universit\'e Lyon 1, Villeurbanne CNRS/IN2P3, UMR5822, Institut de Physique des 2 Infinis de Lyon}
\affiliation[2]{Department of Physics, University of Johannesburg, PO Box 524, Auckland Park 2006, South Africa}

\emailAdd{g.cacciapaglia@ipnl.in2p3.fr}
\emailAdd{acornell@uj.ac.za} 
\emailAdd{cot@ipnl.in2p3.fr}
\emailAdd{deandrea@ipnl.in2p3.fr}
\abstract
{We present a minimal model of asymptotic grand unification based on an $SU(5)$ gauge theory in a compact $S^1/(\mathbb{Z}_2 \times \mathbb{Z}'_2)$ orbifold. The gauge couplings run to a unified fixed point in the UV, without supersymmetry. By construction, fermions are embedded in different $SU(5)$ bulk fields. As a consequence, baryon number is conserved, thus preventing proton decay, and the lightest Kaluza-Klein tier consists of new states that cannot decay into standard model ones. We show that the Yukawa couplings can be either in the bulk or localized, and run to an asymptotically free fixed point in the UV. The lightest massive state can play the role of Dark Matter, produced via baryogenesis, for a Kaluza-Klein mass of about $2.4$~TeV.}

\begin{document}

\hspace*{112mm}{\large \tt } \\
\maketitle
\flushbottom

%%%%%%%%%%%%%%%%%%%%%%%%%%%%%%%%%%%%%%%%%%%%
%  Section 1: Introduction

\section{Introduction}
\label{sec:1}

\par The idea of unification has been employed several times in particle physics when seeking order in the zoo of particles and their interactions. The first famous example is the unification of electromagnetism and the weak force within the semi-simple gauge group $SU(2)_L \times U(1)_Y$ \cite{Glashow:1961tr}, which then became the core of the Standard Model (SM) \cite{Weinberg:1967tq,Salam:1968rm}. Later, quarks and leptons were unified within the Pati-Salam gauge symmetry $SU(4)\times SU(2)_L \times SU(2)_R$ \cite{Pati:1974yy}, which also features a left-right symmetric structure. Ultimately, Grand Unified Theories (GUTs) have the ambition of unifying all the forces (except gravity) into a unique simple gauge group (see, for example, Refs.~\cite{Georgi:1974sy,Georgi:1974yf,Ross:1985ai,Mohapatra:1986uf}). The first realistic example was provided by the Georgi-Glashow theory \cite{Georgi:1974sy}, based on an $SU(5)$ gauge symmetry. In fact, the most minimal chiral set of fields can consist of a $\bf 10$ and a $\bf \bar{5}$ representations, which allows to accommodate a single complete family of the SM, without any gauge anomaly. 

\par It is usual, in GUT model building, to assume that the unification of gauge couplings occurs at a specific high scale, where the low energy couplings meet via the renormalization group running. In the SM, exact unification does not occur, as the gauge couplings tend to similar values in a region of energies around $10^{13}$ to $10^{16}$~GeV. It is well known that exact unification can only be achieved by enlarging the content of the SM fields, usually thanks to supersymmetric extensions \cite{Dimopoulos:1981zb,Amaldi:1991cn} or by adding heavy fields \cite{Amaldi:1991zx,Kopp:2009xt}. What drives unification is the presence of incomplete multiplets of the GUT gauge symmetry, which contributes to the differential running among the three SM gauge couplings \cite{Amaldi:1991zx}.

\par In this work we consider a different kind of unification, one where the couplings unify asymptotically. Thus, instead of crossing at a fixed energy scale, they tend to the same value in the deep Ultra-Violet (UV). Contrary to the standard lore, asymptotic unification is driven by the contributions of complete multiplets of the GUT symmetry.  This requirement is met in models where a compact extra dimension becomes relevant at scales higher than the electroweak (EW) scale and where the gauge symmetry in the bulk is unified. The contribution of the Kaluza-Klein states to the running, therefore, drives the gauge couplings to a unified value at high energies \cite{Dienes:2002bg}. While the four-dimensional gauge couplings run to zero, the five-dimensional gauge couplings run towards a fixed point in the UV \cite{Gies_2003}. This mechanism has been used to define renormalizable theories in five dimensions (5D) \cite{Morris:2004mg} and gauge-Higgs unification models \cite{Khojali:2017azj}. The asymptotic unification is more natural than the more traditional one in extra dimensional models: in fact, at energies above the inverse radius of the compact dimension, the theory approaches genuinely 5D dynamics, where a single gauge coupling is present. The deeper in the UV we probe the theory, the more closely it will behave like a unified theory. Extra-dimensional GUTs \cite{Kawamura:1999nj,Kawamura:2000ir,Contino:2001si,Hall:2001xb}, in the traditional sense, have been considered both in 5D \cite{Kawamura:2000ev,Haba:2001ci,Altarelli:2001qj,Hall:2001pg,Nomura:2001mf,Hebecker:2001wq,Li:2001qs,Haba:2001sy,Dermisek:2001hp} and 6D \cite{Asaka:2001eh,Hall:2001xr,Watari:2001pj,Haba:2001ze,Haba:2002ek}, often accompanied by supersymmetry to achieve exact unification, while the idea of asymptotic unification has only been pioneered in Ref.~\cite{Dienes:2002bg}, without a detailed model being put forward.
Asymptotic unification may also occur in asymptotically safe models at large $N_f$ \cite{Antipin:2017ebo}, where the gauge couplings tend to the same interactive UV fixed point. Explicit examples can be found in Refs \cite{Molinaro:2018kjz,Sannino:2019sch}, where a preliminary standard unification to the Pati-Salam gauge group is imposed to tame the Abelian gauge \cite{Antipin:2018zdg}.

\par Here, we present the first \emph{asymptotic GUT (aGUT)} based on an $SU(5)$ model in a flat $S^1/\mathbb{Z}_2 \times \mathbb{Z}'_2$ orbifold. The choice of the $SU(5)$ group structure stems from the Georgi-Glashow model \cite{Georgi:1974sy}, where we have the smallest simple Lie group that contains the SM one. Historically, this group structure allowed for a reinterpretation of several of the fields as being different states of a single multiplet. The enticing structure of all known matter fields (fermions) fitting perfectly into three copies of the smallest group representations of $SU(5)$, and having the correct quantized charges, became a crucial reason for people to believe that GUTs may be realized in Nature. We recall that the $\bf \bar{5}$ representation contains the charge-conjugate of the right-handed down-type quark and the left-handed lepton iso-spin doublet, while the $\bf 10$ contains the left-handed down-type quark iso-spin doublet, the charge-conjugate of the right-handed up-type quark and of the charged lepton. Thus, a single SM generation can be introduced in an anomaly-free way as a $\bf \bar{5} + 10$. In our extra-dimensional $SU(5)$ model, the chiral SM fermions are identified with the zero modes of the bulk fields. As such, it is not possible to include one whole SM generation in a single set of $SU(5)$ representations. This is due to the orbifold parity that breaks the gauge symmetry. Thus, for each SM family, we need to introduce a set of bulk fields in the representations $\bf 5$, $\bf \bar 5$, $\bf{10}$ and $\bf \bar{10}$. The SM zero modes, therefore, arise from different $SU(5)$ multiplets. The main consequence of this set-up is that baryon number is preserved, thus avoiding the strong constraints on the GUT scale from proton decay, similar to the model in Ref.~\cite{Fornal_2017}. 

\par The lightest Kaluza-Klein (KK) tier is made of new particles that have non-SM quantum numbers, which we baptize \emph{Indalo}-states, from the Zulu word ``indalo", meaning creation (or nature). Curiously, the same name is shared by the symbol \Indalo\ found in prehistoric caves near Almer\'{i}a in Andalusia, Spain. As we shall see in the following, the \Indalo -particles are neither leptons nor quarks and have interesting properties forbidding, for example, proton decay and their decay to a purely SM final state. As the $SU(5)$ gauge symmetry is broken mainly via the orbifold projection, the scalar sector simply consists of a $\bf 5$ scalar multiplet, containing the Higgs doublet and its \Indalo -partner (a QCD color triplet Higgs $H$). By construction, the new \Indalo -particles have unusual baryon and lepton numbers, which forbid their decays into pure SM final states. Thus, the lightest \Indalo -state, corresponding to a neutrino partner, is stable and may constitute the Dark Matter relic density.

\par Neutrinos are also an important part of any model building beyond the SM, thus we shall also consider the effect of adding either Dirac or Majorana neutrinos to this model. Recall that in the traditional $SU(5)$ GUT models, the right-handed neutrinos are singlets of $SU(5)$, which implies that their mass is not forbidden by any symmetry.

\par The manuscript is organized as follows: In section~\ref{sec:2} we introduce the model and the orbifold projections giving rise to the desired spectrum. We also show how baryon (and lepton) number conservation arises. In section~\ref{sec:3} we demonstrate the asymptotic unification of the SM gauge couplings using the renormalization group equations. Furthermore, in section~\ref{sec:4} we discuss the running of the Yukawa couplings and discuss the implications for flavor bounds and neutrino mass generation. In section~\ref{sec:5} we discuss phenomenological aspects of the model, in particular baryogenesis and Indalogenesis as a potential Dark Matter generation mechanism.  Finally, we offer our conclusions in section~\ref{sec:7}.

%%%%%%%%%%%%%%%%%%%%%%%%%%%%%%%%%%%%%%%%%%%%
%  Section 2: The Model

\section{The Model}\label{sec:2}

\par We consider here the minimal aGUT model, as outlined in the introduction. It is based on an $SU(5)$ gauge symmetry in the bulk, broken down to the SM gauge symmetry, $SU(3)_c$ $\times$ $SU(2)_L$ $\times$ $U(1)_Y$, via orbifold boundary conditions. The background corresponds to a single extra-dimension compactified on an orbifold $S^1/(\mathbb{Z}_2 \times \mathbb{Z}_2')$ of radius $R$. The gauge breaking is achieved by use of two parities, defined by two matrices $P_0$ and $P_1$ in the gauge space, which correspond to a mirror symmetry around the fixed points $y = 0$ and $y = \pi R/2$ respectively. They are chosen such that $SU(5)$ is broken by $P_0$ at $y=0$ while being preserved on the boundary at $y = \pi R/2$. Whilst this is not a unique construction to achieve the aGUT scenario, it is the simplest and most minimal realization. To further develop this model we shall first explore the orbifold parities in section~\ref{sec:2-1}, before introducing the Yukawa sector of our model and, finally, the complete Lagrangian in section~\ref{sec:2-2}.

%%%%%%%%%%%%%%%%%%%%%%%%%%%%%%%%%%%%%%%%%%%%
\subsection{Boundary conditions}\label{sec:2-1}

\par The two parities acting on the fields are associated to two matrices in the gauge space, $P_0$ and $P_1$, corresponding to the two fixed points $y = 0$ and $y = \pi R / 2$ respectively. For the gauge fields $A^a_M$, where $M$ is the 5D Lorentz index, this implies the following relations \cite{Haba:2002py}:
\begin{equation}
    (P_0) \Rightarrow \left\{ \begin{array}{l} A^a_{\mu}(x,-y)  = P_0 A^a_{\mu}(x,y) P_0^{\dagger}\,,
\\
    A^a_y(x,-y) = - P_0 A^a_y(x,y) P_0^{\dagger} \,,
\end{array} \right.
\end{equation}
\begin{equation}
    (P_1) \Rightarrow \left\{ \begin{array}{l} A^a_{\mu}(x,\pi R - y) = P_1 A^a_{\mu}(x, y) P_1^{\dagger}\,,
    \\
    A^a_y(x,\pi R - y) = - P_1 A^a_y(x, y) P_1^{\dagger} \,,
\end{array} \right.
\end{equation}
where $A_y$ is the polarization along the fifth compact dimension, and the fields are periodic over $y \to y + 2 \pi R$. The radius $R$ defines the mass scale of the KK modes of each field. In order to have the desired boundary conditions we choose the $P_0$ and $P_1$ matrices to be diagonal with elements:
\begin{eqnarray}
P_0 &=& \begin{pmatrix} + & + & + & - & -
\end{pmatrix} \, , \\
P_1 &=& \begin{pmatrix} + & + & + & + & +
\end{pmatrix} \, .
\end{eqnarray}
With this choice the bulk $SU(5)$ is broken to the SM group on the $y=0$ boundary. We also recall that parity-odd fields respect Dirichlet boundary conditions (i.e. vanishing of the field), while parity-even ones respect Neumann boundary conditions (i.e. vanishing of the derivative of the field). Thus, all the gauge-scalar modes are ``eaten" by the massive vector KK modes. The mode decomposition is summarized in Table~\ref{tab:1}.

\par The Higgs boson is embedded in a bulk scalar field $\phi_5$, transforming as a fundamental $\bf 5$ of $SU(5)$, with parities:
\begin{equation}\label{eqn:2-5}
    (P_0) \Rightarrow \phi_5(x,-y) = - P_0 \phi_5(x,y) \, ,
\end{equation}
\begin{equation}\label{eqn:2-6}
    (P_1) \Rightarrow \phi_5(x,\pi R - y) = + P_1 \phi_5(x, y) \, .
\end{equation}
As in standard $SU(5)$ GUT models, the Higgs, $\phi_h$, is accompanied by a QCD-triplet scalar $H$:
\begin{equation}
    \phi_5 = \begin{pmatrix} H \\ \phi_h
\end{pmatrix} \, ,
\end{equation}
where $H$ has no zero mode due to the Dirichlet boundary conditions on the $y=0$ boundary.

\par In $SU(5)$ GUTs, one SM fermion generation is usually embedded into a chiral set of $\bf \bar{5} + 10$. In the 5D construction, however, due to the boundary conditions on $y=0$ that breaks $SU(5)$, it is not possible to embed a complete set of SM fermions with zero modes in a single field\footnote{In Ref.~\cite{Blasi:2020ktl} the possibility of using universal boundary conditions had been used in the context of holographic composite Higgs models. This choice, however, goes beyond the orbifold construction we use here.}. Therefore our model set-up requires a doubling of the number of bulk fields as follows: 
\begin{equation}
\psi_{1_{L/R}} = N\,, \quad
    \psi_{5_{L/R}} = \begin{pmatrix} b \\ L^c
\end{pmatrix}_{L/R} \, , \quad
    \psi_{\overline{5}_{L/R}} = \begin{pmatrix} B^c \\ l
\end{pmatrix}_{L/R} \,,
\end{equation}
\begin{equation}
\psi_{10_{L/R}} = \frac{1}{\sqrt{2}}\begin{pmatrix} T^c & q \\ & \mathcal{T}^c
\end{pmatrix}_{L/R} \,, \quad
\psi_{\overline{10}_{L/R}} = \frac{1}{\sqrt{2}}\begin{pmatrix} t & Q^c \\ & \tau
\end{pmatrix}_{L/R} \, ,
\end{equation}
where the capitalized letters indicate fields that do not have a zero mode, and the superscript ${}^c$ indicates the 4D charge conjugate. We also included a singlet $\psi_1$ in each generation, to play the role of the right-handed neutrino and generate neutrino masses. The parities are chosen as follows:
\begin{equation}\label{eqn:2-8}
    (P_0) \Rightarrow \left\{ \begin{array}{l} \psi_1(x,-y) = - \gamma_5 \psi_1(x,y) \,,
    \\\psi_5(x,-y) = + P_0 \gamma_5 \psi_5(x,y) \,,
    \\
    \psi_{\overline{5}}(x,-y) = + P_0^{\dagger} \gamma_5 \psi_{\overline{5}}(x,y) \,, \\
    \psi_{10}(x,-y) = + P_0 \gamma_5 \psi_{10}(x,y) P_0^{T} \, , 
    \\
    \psi_{\overline{10}}(x,-y) = + P_0^{\dagger} \gamma_5 \psi_{\overline{10}}(x,y) P_0^{*} \,; 
\end{array} \right.
\end{equation}
\begin{equation}\label{eqn:2-9}
    (P_1) \Rightarrow \left\{ \begin{array}{l} \psi_1(x,\pi R - y) = - \gamma_5 \psi_1(x, y) \,,
    \\\psi_5(x,\pi R - y) = + P_1 \gamma_5 \psi_5(x, y) \,,
    \\
    \psi_{\overline{5}}(x,\pi R - y) = - P_1^{\dagger} \gamma_5 \psi_{\overline{5}}(x, y) \,, \\
    \psi_{10}(x,\pi R - y) = - P_1 \gamma_5 \psi_{10}(x, y)  P_1^{T} \,,
    \\
    \psi_{\overline{10}}(x,\pi R - y) = + P_1^{\dagger} \gamma_5 \psi_{\overline{10}}(x, y)  P_1^{*} \, .
\end{array} \right.
\end{equation}
In the above equations $\gamma_5$ is the Dirac matrix in the (diagonal) Weyl representation. The quantum numbers of all the components are reported in Table~\ref{tab:1}.
\begin{table}[ht]
\begin{center}
\begin{tabular}{ |l|c|c|c|c|}
\hline
Field & $(\mathbb{Z}_2, \mathbb{Z}'_2)$ & SM & zero mode? & KK mass \\ \hline
$l$ & $(+,+)$ & \multirow{2}{*}{$({\bf 1}, {\bf 2}, -1/2)$} & $\surd$ & $2/R$ \\
$L$ & $(+,-)$ &  & $-$ & $1/R$ \\ 
\hline
$\tau$ & $(-,-)$ & \multirow{2}{*}{$({\bf 1}, {\bf 1}, -1)$} & $\surd$ & $2/R$ \\
$\mathcal{T}$ & $(-,+)$ &  & $-$ & $1/R$ \\ 
\hline
$N$ & $(-,-)$ & $({\bf 1}, {\bf 1}, 0)$ & $\surd$ & $2/R$ \\
\hline
$q$ & $(+,+)$ & \multirow{2}{*}{$({\bf 3}, {\bf 2}, 1/6)$} & $\surd$ & $2/R$ \\
$Q$ & $(+,-)$ &  & $-$ & $1/R$ \\ 
\hline
$t$ & $(-,-)$ & \multirow{2}{*}{$({\bf 3}, {\bf 1}, 2/3)$} & $\surd$ & $2/R$ \\
$T$ & $(-,+)$ &  & $-$ & $1/R$ \\ 
\hline
$b$ & $(-,-)$ & \multirow{2}{*}{$({\bf 3}, {\bf 1}, -1/3)$} & $\surd$ & $2/R$ \\
$B$ & $(-,+)$ &  & $-$ & $1/R$ \\ 
\hline
$\phi_h$ & $(+,+)$ & $({\bf 1}, {\bf 2}, 1/2)$ & $\surd$ & $2/R$ \\
$H$ & $(-,+)$ & $({\bf 3}, {\bf 1}, -1/3)$ & $-$ & $1/R$ \\ 
\hline
$B_\mu$ & \multirow{3}{*}{$(+,+)$} & $({\bf 1}, {\bf 1}, 0)$ & \multirow{3}{*}{$\surd$} & \multirow{3}{*}{$2/R$} \\
$W_\mu^a$ && $({\bf 1}, {\bf 3}, 0)$ & & \\
$G_\mu^i$ && $({\bf 8}, {\bf 1}, 0)$ & & \\
\hline
$A_X^\mu$ & $(-,+)$ & $({\bf 3}, {\bf 2}, -5/6)$ & $-$ & $1/R$ \\ 
\hline
\end{tabular}\end{center}
\caption{Quantum numbers and parities of all the 5D fields (for fermions we indicate the parities of the left-handed chiralities). The last two columns indicate the presence of a zero mode, and the mass of the lightest non-zero KK mode.}\label{tab:1}
\end{table}

\par The above-mentioned fields in five dimensions can be decomposed into towers of KK modes, whose characteristics depend on the parities under the $\mathbb{Z}_2$ and $\mathbb{Z}'_2$, which we denote $(\pm, \pm)$ (for fermions we always denote the parities of the left-handed chirality). Of the four combinations only $(+,+)$ features a zero mode ($(-,-)$ for the right-handed chirality of the fermions), which can be associated to a SM field. In the last two columns of Table \ref{tab:1} we indicate the presence of a zero mode and the mass of the lightest non-zero KK mode for each field.

\par The lightest tier of non-zero KK modes, with mass $m_{\rm KK} = 1/R$, is populated by the fields that do not have a zero mode: a complete copy of the SM fermion families, the QCD-charged Higgs $H$ and the $SU(5)$ vector lepto-quarks $A_X^\mu$. As we will see at the end of the section, these fields play a special role in this model, thus we collectively name them ``Indalo" (\Indalo -states). For now, only the right-handed neutrino has no Indalo partner, even though such an Indalo-neutrino will be introduced later for phenomenological reasons.

%%%%%%%%%%%%%%%%%%%%%%%%%%%%%%%%%%%%%%%%%%%%
\subsection{Fermion non-unification}\label{sec:2-2}

\par As we have seen above, due to the $SU(5)$ breaking parity at the $y=0$ boundary, it is not possible to embed all the SM fermions (via zero modes) in the same bulk fermion field. Thus we introduced two copies, with opposite quantum numbers and opposite parities. As such, this model naturally produces a ``fake GUT'' \cite{Ibe:2019ifm} structure in the fermion sector. An important consequence of this property is that the Yukawa couplings are not required to unify.

\par With the set of fields introduced above, we can write down the most general bulk Lagrangian as follows:
\begin{eqnarray}
 \mathcal{L}_{SU(5)} &= & -\frac{1}{4}F^{(a)}_{M N}{F^{(a)}}^{M N} - 
 \frac{1}{2\xi}(\partial_{\mu}A^{\mu} - \xi \partial_{5}A_y)^2 + i \overline{\psi_{5}}\slashed{D}\psi_5 + i \overline{\psi_{\overline{5}}}\slashed{D}\psi_{\overline{5}} + i \overline{\psi_{10}}\slashed{D}\psi_{10}  \notag \\
&+& i \overline{\psi_{\overline{10}}}\slashed{D}\psi_{\overline{10}} - \left( \sqrt{2} Y_{\tau}\,  \overline{\psi_{\overline{5}}} \psi_{\overline{10}}\phi_5^* + \sqrt{2} Y_{b}\,  \overline{\psi_{5}} \psi_{10}\phi_5^* + \frac{1}{2} Y_{t}\,  \epsilon_5\ \overline{\psi_{\overline{10}}} \psi_{10}\phi_5 + \mbox{h.c.}\right)  
 \notag \\
 &+& |D_M \phi_5|^2 - V (\phi_5) + i \overline{\psi_1}\slashed{\partial}\psi_{1} - \left( Y_\nu\, \overline{\psi_1} \psi_{\bar{5}} \phi_5 + \mbox{h.c.} \right)\,, \label{Model-Lagrangian}
\end{eqnarray}
where $D_M$ = $\partial_M$ - $i g^a T^a A^a_M$ with $T^a$ being the $SU(5)$ generators in the appropriate representation, and $\epsilon_5$ is the 5-dimensional Levi-Civita symbol on the gauge indices. The term $V(\phi_5)$ represents a generic potential for the scalar field, which is responsible for generating a vacuum expectation value for the Higgs zero mode, like in the SM. We also recall that gauge and Yukawa couplings, via naive dimensional analysis in 5D, have scaling dimension $[m]^{-1/2}$. 

\par The normalization of the Yukawa couplings is chosen to reproduce the SM ones for the zero modes. Expanding the $SU(5)$ multiplets into their components, the Yukawa couplings contain the following terms: 
\begin{eqnarray}
\overline{\psi_1} \psi_{\bar{5}} \phi_5 &=& \overline{N} \phi_h l + \overline{N} H \mathcal{B}^c \,, \\
\sqrt{2}\ \overline{\psi}_{5} \psi_{10}\phi_5^* &=&  \overline{b}\phi_h^{*}q - \overline{L}^cH^*q - \overline{L}^c\phi_h^*\mathcal{T}^c + \epsilon_3 \overline{b}H^{*}T^{c}\,, \\ \sqrt{2}\ \overline{\psi_{\overline{10}}} \psi_{\overline{5}}\phi_5^* &=&  - \overline{\tau}\phi_h^*l - \overline{Q}^{c}H^{*}l + \overline{Q}^{c}\phi_h^{*}B^{c} - \epsilon_3 \overline{t}H^{*}B^{c}\,, \\
\frac{1}{2} \ \epsilon_5\overline{\psi_{\overline{10}}} \psi_{10}\phi_5 & = & \overline{t}\phi_h q + \overline{t}H\mathcal{T}^c + \overline{\tau}HT^{c} + \overline{Q^{c}}\phi_hT^{c} + \epsilon_3 \overline{Q}^{c}Hq \,,
\end{eqnarray}
where $\epsilon_3$ is the Levi-Civita symbol contracting the QCD $SU(3)$ indices. We can see from the above equations that each Yukawa coupling contains only one SM-like coupling between zero modes. Once normalized to dimensionless 4D couplings, $y_f = Y_f/\sqrt{2 \pi R}$, they correspond directly to the SM Yukawa couplings, and can be extended to the full flavor structure. 

%%%%%%%%%%%%%%%%%%%%%%%%%%%%%%%%%%%%%%%%%%%%
\subsection{Baryon and Lepton number conservation}\label{sec:2-3}

\par In $SU(5)$ models, strong constraints usually arise from proton decay considerations, via baryon violating couplings of the QCD-charged scalar $H$ \cite{Georgi:1974sy,Golowich:1981sb}. However, the violation of baryon and lepton number occurs because the SM fields are embedded in the same $SU(5)$ multiplet, while in our model they are not by construction, due to the specific structure in 5D. 

\par From the Lagrangian in Eq.~\eqref{Model-Lagrangian} we see that we can assign two independent global charges to the five fermion and one scalar fields (due to the presence of 4 independent Yukawa couplings). Combined with the hypercharge, which is embedded in $SU(5)$, we find two linear combinations of the three $U(1)$'s that are preserved by the Higgs vacuum expectation value. Their assignments on the multiplet components are listed in Table~\ref{tab:2}.

\par For the SM fields, containing the zero modes, we can choose the charges to match the standard baryon ($B$) and lepton ($L$) numbers. This ensures that no proton decay is allowed in our aGUT model, and the unification (compactification) scale can be potentially placed at a low scale compared to the standard GUTs. Furthermore, all \Indalo -states carry both $B$ and $L$ charges, in values that are half of the SM unit charges. This implies that it is not possible for them to decay into SM fields only. This property makes the lightest \Indalo -state stable, and potentially a candidate for Dark Matter, as we will explore in section~\ref{sec:5}.
\begin{table}[ht]
\begin{center}
\begin{tabular}{ |p{1.6cm}||p{1cm}||p{1cm}|p{1cm}|p{1cm}|p{1cm}| }
 \hline
 Multiplets    & Fields  & L & B & Q & $Q_3$\\
 \hline
 $\psi_{\overline{5}}$ & $B_R^c$ & 1/2 & 1/6 & 1/3 & 0\\
  &  \fbox{$\tau_L$}  & 1 & 0 & -1 & -1\\
  &  \fbox{$\nu_L$}  & 1 & 0 & 0 & 1\\
 \hline
 $\psi_{5}$ & \fbox{$b_R$} &  0 & 1/3 & -1/3 & 0 \\
 &  $\mathcal{T}_L^c$  & -1/2 & 1/2 & 1 & 1\\
 &  $\mathcal{N}_L^c$  & -1/2 & 1/2 & 0 & -1\\
 \hline
  $\psi_{10}$ & $T_R^c$ &  1/2 & 1/6 & -2/3 & 0\\
  & $\mathcal{T}_R^c$ & -1/2 & 1/2 & 1 & 0\\
  & \fbox{$t_L$} &  0 & 1/3 & 2/3 & 1\\
  & \fbox{$b_L$} &  0 & 1/3 & -1/3 & -1\\
  \hline
  $\psi_{\overline{10}}$ & \fbox{$t_R$} & 0 & 1/3 & 2/3 & 0\\
  & \fbox{$\tau_R$}  & 1 & 0 & -1 & 0\\
  & $T_L^c$ & 1/2 & 1/6 & -2/3 & -1 \\
  & $B_L^c$ & 1/2 & 1/6 & 1/3 & 1 \\
 \hline
 $\psi_1$ & $N$ & $1$ & $0$ & $0$ & $0$ \\ 
 \hline
 $\phi_5$ & $H$ & 1/2 & -1/6 & -1/3 & 0\\
 & \fbox{$\phi^+$} & 0 & 0 & 1 & 1\\
 & \fbox{$\phi_0$} & 0 & 0 & 0 & -1\\
 \hline
  $A_X$ & $X$ & 1/2 & -1/6 & -4/3 & -1\\
 & $Y$ & 1/2 & -1/6 & -1/3 & 1\\
 \hline
\end{tabular} \end{center}
\caption{Baryon and lepton numbers for the components of the $SU(5)$ multiplets. We also indicate their electromagnetic charge $Q$ and weak iso-spin $Q_3$.}\label{tab:2}
\end{table}

%%%%%%%%%%%%%%%%%%%%%%%%%%%%%%%%%%%%%%%%%%%%
%  Section 3: Asymptotic running and unification

\section{Gauge running and asymptotic unification}\label{sec:3}

\par In traditional GUT model building the gauge couplings are supposed to run up to the same value at a given high scale. Thus, the most relevant feature of the renormalization group equations (RGEs) is the relative evolution of the couplings. This is provided by the gauge bosons and the Higgs, which come in incomplete multiplets of the GUT $SU(5)$ symmetry \cite{Dimopoulos:1981xm}. The same approach has been considered in GUT models in extra dimensions \cite{Kawamura:2000ev,Haba:2001ci,Contino:2001si,Hall:2001xb}.

\par Here we will consider an antithetic scenario where it is the $SU(5)$--invariant running, provided by the bulk KK modes, that drives unification of the gauge couplings \emph{asymptotically} at high scales. This possibility was first noted in Ref.~\cite{Dienes:2002bg}, and applied to gauge-Higgs unification models in Refs~\cite{Morris:2004mg,Khojali:2017azj,Khojali:2017rnl,Khojali:2017ejm}. Note that the RGEs for the gauge couplings are SM-like up to the compactification scale $1/R$, where the effect of the KK states enters. As in standard $SU(5)$ models, we will follow the evolution of the $SU(3)\times SU(2) \times U(1)$ couplings $g_i$ = $\{g_1,g_2,g_3\}$, where the hypercharge coupling is normalized as $g_1 = \sqrt{\frac{5}{3}}g'$. The RGEs can be written as \cite{Bhattacharyya:2006ym}:
\begin{equation}
    2 \pi \frac{d\alpha_i}{dt} = b_i^{\rm SM} \alpha_i^2 + (S(t) - 1)\  b_{5} \alpha_i^2 \,,
\end{equation}
where $\alpha_i = g_i^2/4\pi$, $t = \mathrm{ln}(\mu / m_Z)$, $m_Z$ is the $Z$ boson mass, and the SM coefficients read $(b_1^{\rm SM}, b_2^{\rm SM}, b_{3}^{\rm SM}) = (41/10, -19/6,-7)$. The second term includes the contributions of the KK states, in a continuum approximation, contained in the function
\begin{equation}
    S(t) = \left\{
    \begin{array}{ll}
        \mu R = m_Z R e^t & \mbox{for } \mu > 1/R\,,   \\
        1 & \mbox{for } m_Z < \mu < 1/R \,.
    \end{array}
\right.
\end{equation}
As the KK modes tend to appear in complete multiplets of $SU(5)$ (more precisely, adding the KK states of adjacent even and odd KK tiers), all gauge couplings receive the same beta function, given by
\begin{equation}
 b_5 = - \frac{52}{3} + \frac{16}{3} n_{\rm g}\,,   \label{eq:b5}
\end{equation}
where $n_{\rm g}$ is the number of fermion generations in the bulk. For 3 families, as we will consider in the following, we find $b^{SU(5)} = - 4/3$. Note that this value allows for the 5D theory to have an UV fixed point \cite{Gies_2003,Morris:2004mg}, a fact that will play a crucial role for the asymptotic unification.

\par
\begin{figure}[ht!]
\begin{centering}
\includegraphics[scale=1]{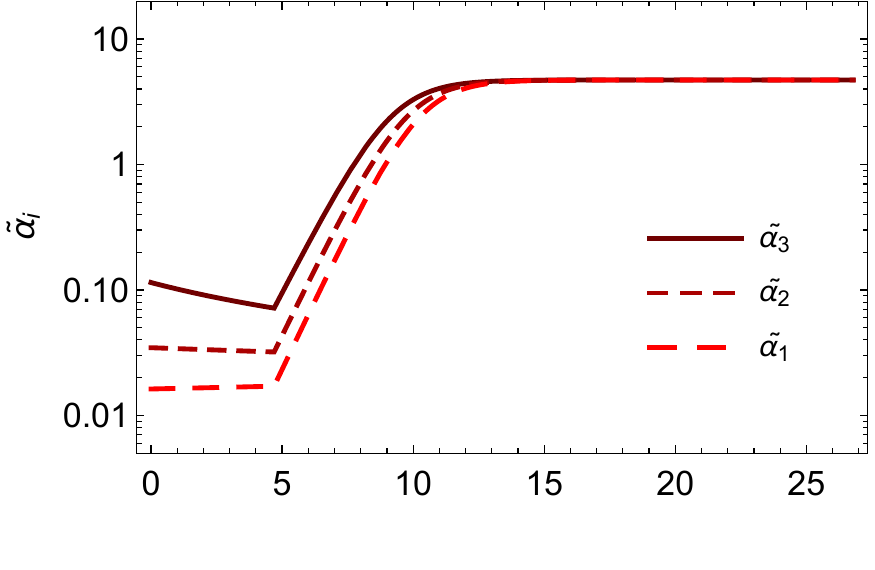}
\caption{Running of the gauge couplings using one-loop factors, with $R^{-1} = 10$~TeV. The range of $t$ corresponds to the $Z$ mass ($t=0$) and the reduced 5D Planck mass.}\label{fig:1}
\end{centering}
\end{figure}

 Using the RGEs for 3 families of fermions we show in Fig.~\ref{fig:1} the one-loop evolution of the three gauge couplings in terms of 
\begin{equation}
\tilde{\alpha}_i = \left\{ \begin{array}{l} \alpha_i(t)\;\; \mbox{for}\;\; \mu < 1/R\,, \\
\alpha_i(t) S(t)\;\; \mbox{for}\;\; \mu > 1/R\,.
\end{array} \right. \label{eq:alphatilde}
\end{equation}
At energies above the compactification scale we consider an effective 't Hooft coupling, which takes into account the number of KK modes below the energy scale $\mu$. We start the running at the $Z$ mass with the SM values $\{g_1^0, g_2^0, g_3^0\}$ = $\{0.45,0.66,1.2\}$, while the matching to the 5D running takes place at the scale $1/R$, indicated by the point where the running changes sharply. Because of the absence of baryon and lepton number violation, this scale can be low, so we choose $1/R = 10$~TeV as a benchmark in the plot. We can see that the couplings never cross, however, they do get very close and tend to a unified value asymptotically at high energies. In fact, this value corresponds to the UV safe fixed point of the 5D theory. At $t \approx 10$ the couplings are effectively unified. This scale is well below the 5D reduced Planck mass $M_{\rm Pl}^\ast$ \cite{ArkaniHamed:1998nn}, which corresponds to the largest value of $t$ shown in the plot. Increasing $1/R$ does not change the picture qualitatively.

\par The asymptotic behavior of the gauge couplings can be understood once the RGEs are rewritten in terms of $\tilde \alpha_i$ at large energies. Keeping the leading terms in $1/\mu R$, the RGE is the same for all gauge couplings:
\begin{equation}
2 \pi \frac{d\tilde\alpha_i}{dt} = 2 \pi \tilde\alpha_i + b_5\ \tilde\alpha_i^2\,. 
\end{equation}
The beta function vanishes at 
\begin{equation}
\tilde\alpha_i^\ast (IR)= 0\,, \quad \tilde\alpha_i^\ast (UV) =  -\frac{2 \pi}{b_5}\,,
\end{equation}
which are the IR and the UV fixed points, respectively. The UV fixed point only exists for $b_5 < 0$, thus using the result in Eq.~\eqref{eq:b5} the fixed point's existence requires $n_{\rm g} \leq 3$. For $4$ or more bulk generations the asymptotic unification would fail.

\par For 3 bulk generations we find
\begin{equation}
\tilde\alpha_i^\ast = \frac{3 \pi}{2}\,.
\end{equation}
This value of the coupling is apparently non-perturbative. However, a more realistic assessment requires the estimation of the extra-dimensional loop factor, as follows:
\begin{equation}
\xi (d) = \frac{\Omega (d)}{(2 \pi)^d} 4 \pi \tilde\alpha\,,
\end{equation}
where $\Omega (d)$ is the $d$-dimensional solid angle. For $d=5$ and the 3 generation fixed point, we find:
\begin{equation}
\xi (5) = \frac{\tilde\alpha^\ast_i}{3 \pi^2} = \frac{1}{2\pi} < 1\,.
\end{equation}
We thus estimate the theory to remain perturbative at high energies, where the full 5D nature is revealed.

%%%%%%%%%%%%%%%%%%%%%%%%%%%%%%%%%%%%%%%%%%%%
\section{Running of the Yukawa couplings}\label{sec:4}

\par As already mentioned, the Yukawa couplings do not unify, as the bulk $SU(5)$ theory has one independent Yukawa for each quark and lepton. Nevertheless, it is important to study their running and explore the possibility that they may also run to an UV fixed point. If not, consistency of the theory requires the absence of Landau poles below the 5D Planck mass.

\par The calculation of the Yukawa running has an added difficulty compared to the gauge one, due to the fact that gauge couplings enter in the beta function at the one loop level. The issue is that this concerns not only the SM gauge couplings, but also the couplings of the massive \Indalo -states $A_X$. In principle, the running of these couplings is not linked to that of the gauge couplings at low energies. To take into account this uncertainty, we compute the contribution of the KK modes under the assumption of a single 5D coupling $g_5$, and assign to it the extreme values $g_5 = g_1$ and $g_5 = g_3$. We will consider the variation between the two as a systematic uncertainty in our results. 

%%%%%%%%%%%%%%%%%%%%%%%%%%%%%%%%%%%%%%%%%%%%
%\subsubsection*{Coefficients in the SM}
\begin{center}
{\it RGE coefficients in the SM}
\end{center}

\par Firstly, in the SM the four Yukawa couplings run according to the following RGEs:
\begin{eqnarray}
    \left. 2\pi\frac{d\alpha_t}{dt} \right|_{\rm SM} &=& \left[ \frac{9}{2}\alpha_t + \frac{3}{2}\alpha_b + \alpha_{\tau} + \alpha_\nu  - \frac{17}{20}\alpha_1   - \frac{9}{4}\alpha_2 - 8\alpha_3 \right]\alpha_t \,, \\
    \left. 2\pi\frac{d\alpha_b}{dt} \right|_{\rm SM}  &=& \left[\frac{9}{2}\alpha_b + \frac{3}{2}\alpha_t + \alpha_{\tau} + \alpha_\nu
    - \frac{1}{4}\alpha_1 - \frac{9}{4}\alpha_2-8\alpha_3\right]\alpha_b \,, \\
    \left. 2\pi\frac{d\alpha_\tau}{dt} \right|_{\rm SM}  &=& \left[\frac{5}{2}\alpha_{\tau} + 3\alpha_t + 3\alpha_b -\frac{1}{2} \alpha_\nu
    - \frac{9}{4} (\alpha_1+\alpha_2)\right]\alpha_{\tau} \,, \\
    \left. 2\pi\frac{d\alpha_\nu}{dt} \right|_{\rm SM}  &=& \left[\frac{5}{2}\alpha_{\nu} + 3\alpha_t + 3\alpha_b -\frac{1}{2} \alpha_\tau
    - \frac{9}{20} \alpha_1 - \frac{9}{4} \alpha_2\right]\alpha_{\nu} \,,
\end{eqnarray}
where $\alpha_f = y_f^2/4\pi$. These equations will be used for the running of the Yukawa couplings between the EW scale and the compactification scale.\footnote{If the right-handed neutrino $N$ has a Majorana mass, the running of the neutrino Yukawa is also valid above this scale.} Here we only consider the third generation, as they have the largest couplings. The results can be extended to three generations in a straightforward way.

%%%%%%%%%%%%%%%%%%%%%%%%%%%%%%%%%%%%%%%%%%%%
%\subsubsection*{Bulk coefficients}
\begin{center}
{\it RGE coefficients in the bulk}
\end{center}

\par As explained above, we compute the coefficients of the gauge contribution to the Yukawa couplings in an $SU(5)$ unified framework, with gauge coupling $g_5$. For the top Yukawa we obtain:
\begin{equation}
    2\pi \frac{d\alpha_t}{dt} = \left. 2\pi\frac{d\alpha_t}{dt} \right|_{\rm SM} + (S(t)-1)
  \left[ 15 \alpha_t + 13 \left(\alpha_b + \alpha_{\tau}\right) + 2 \alpha_\nu - \frac{114}{5} \alpha_5 \right] \alpha_t\, . 
  \label{eq:RGEtopbulk}
\end{equation}
Similarly, for the bottom, tau and neutrino we find:
\begin{eqnarray}
    2\pi \frac{d\alpha_b}{dt} &=& \left. 2\pi\frac{d\alpha_b}{dt} \right|_{\rm SM} + (S(t)-1)
     \left[ 11 \alpha_b + \frac{39}{2} \alpha_t + 8 \alpha_{\tau} + 2 \alpha_\nu- \frac{93}{5} \alpha_5 \right] \alpha_b \,, 
\label{eq:RGEbottombulk} \\
    2\pi \frac{d\alpha_{\tau}}{dt} &=& \left. 2\pi\frac{d\alpha_\tau}{dt} \right|_{\rm SM} + (S(t)-1)
   \left[ 11 \alpha_{\tau} + \frac{39}{2} \alpha_t + 8 \alpha_{b} + \frac{9}{2} \alpha_\nu - \frac{93}{5} \alpha_5 \right] \alpha_{\tau} \,, \label{eq:RGEtaubulk} \\
    2\pi \frac{d\alpha_{\nu}}{dt} &=& \left. 2\pi\frac{d\alpha_\nu}{dt} \right|_{\rm SM} + (S(t)-1)
   \left[ 5 \alpha_{\nu} + 12 \alpha_t + 8 \alpha_{b} + 18 \alpha_\tau - 6 \alpha_5 \right] \alpha_{\nu} \,. \label{eq:RGEneutrinobulk}
\end{eqnarray}
Numerical results will be expressed in terms of $\tilde\alpha_f$, defined as in Eq.~\eqref{eq:alphatilde}.

%%%%%%%%%%%%%%%%%%%%%%%%%%
\subsection{Numerical results for bulk Yukawas}

Firstly, to check for the tantalizing possibility that the Yukawa couplings run to an interactive UV fixed point, like the gauge couplings, we can check the presence of zeros for the beta function at large energies. Like for the gauge, we can express the RGEs in terms of $\tilde{\alpha}_f$ and expand at the leading order in $1/R\mu$. We find:
\begin{eqnarray}
    & \tilde{\alpha}_t^\ast =  \displaystyle -\frac{14 (2777-140 n_g) \pi}{3585 (13-4 n_g)} \,, \quad \tilde{\alpha}_b^\ast = \frac{ (55889-1160 n_g) \pi}{2390 (13-4 n_g)} \,, & \nonumber \\ &\tilde{\alpha}_\tau^\ast =  \displaystyle -\frac{ (42671-1640 n_g) \pi}{2390 (13-4 n_g)} \,, \quad \tilde{\alpha}_\nu^\ast = \frac{336 (176-5 n_g) \pi}{1195 (13-4 n_g)} \,. &
\end{eqnarray}
The fact that the zeroes for the top and tau Yukawas are always negative (for $n_g \leq 3$) implies the absence of a completely safe fixed point.

\begin{figure}[tb!]
\begin{centering}\begin{tabular}{cc}
 \phantom{xxxxxxxx} $\large \bf R^{-1} = 3.05\cdot 10^5~\mbox{TeV}$ & \phantom{xxxxxxxx} $\large \bf R^{-1} = 10^{10}~\mbox{TeV}$ \\ 
\includegraphics[scale=0.78]{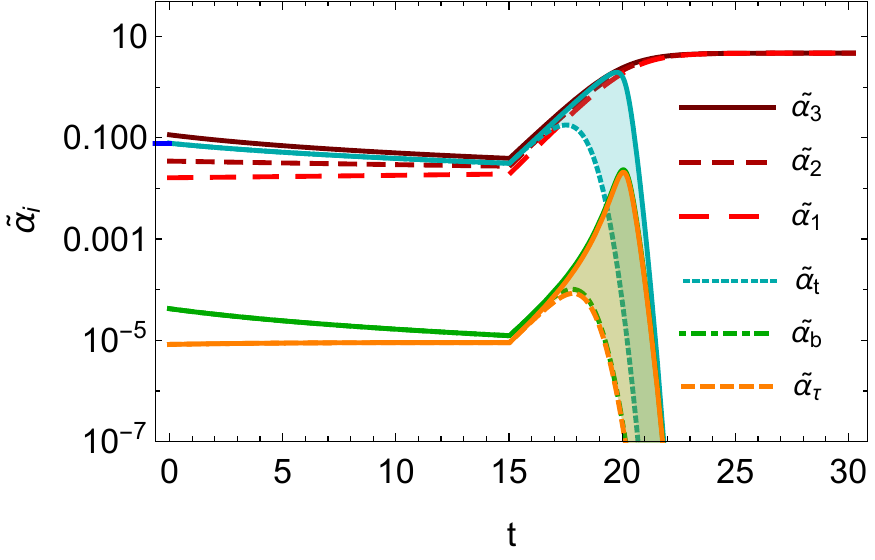} &
    \includegraphics[scale=0.78]{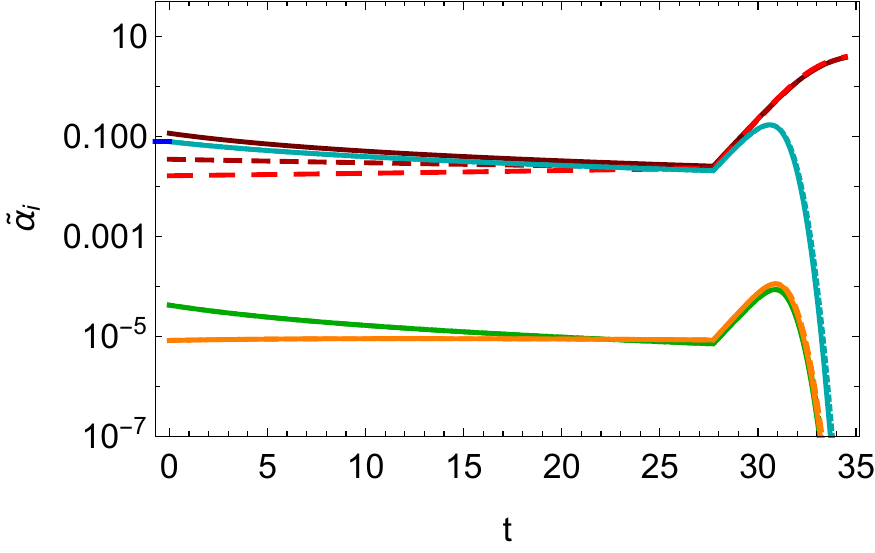} \\
\phantom{xxxxxxxx} $\large \bf R^{-1} = 2.4~\mbox{TeV}$ & \phantom{xxxxxxxx} $\large \bf R^{-1} = 10^{3}~\mbox{TeV}$ \\ \includegraphics[scale=0.78]{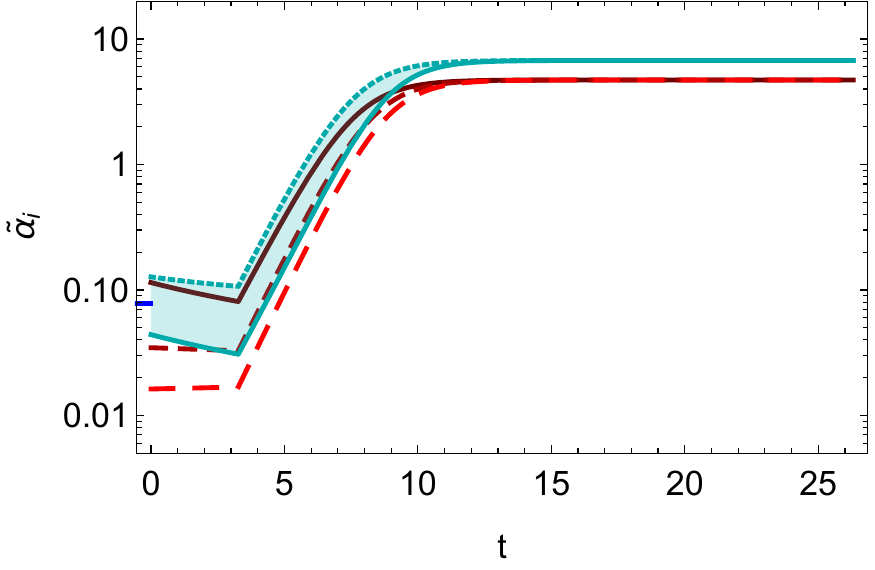} &
    \includegraphics[scale=0.78]{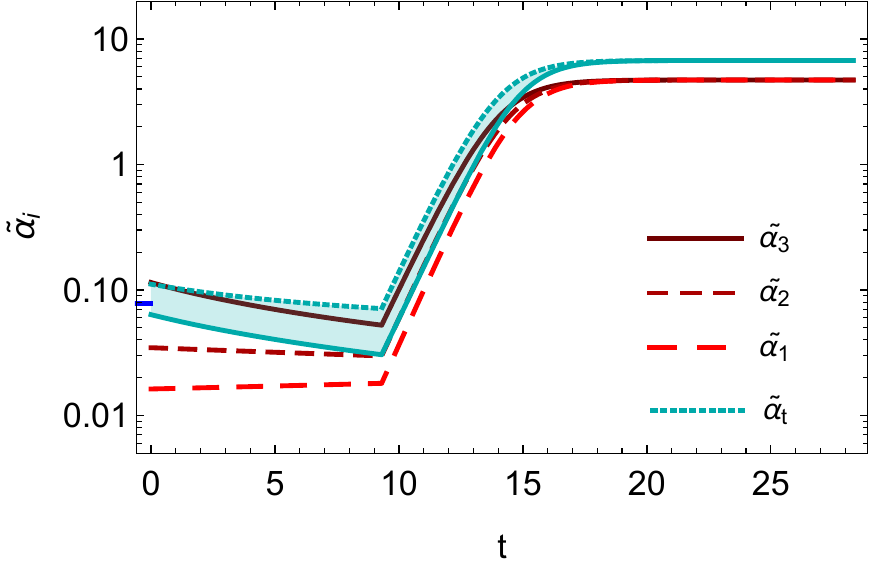} 
\end{tabular} \end{centering}
    \caption{Running of the bulk Yukawas as compared to the gauge couplings. In the top row, we run up from the EW scale for two sample values of compactification scales above the critical value. In the bottom row, we run the top Yukawa down from the UV fixed point (imposed at the 5D Planck scale) for two sample values of the compactification scales below the critical value. The bands indicate the systematic uncertainty from the gauge couplings, while the SM value of the top Yukawa at the EW scale is indicated by the blue tick at $t=0$. The largest value of $t$ in the plots corresponds to the 5D Planck scale.}
    \label{fig:bulk}
\end{figure}

Thus, the only physical possibility to keep all Yukawa couplings in the bulk is that they may run to zero in the UV. Numerically, we found that for
\begin{equation} \label{eq:limrun}
    \frac{1}{R} \gtrsim 3 \cdot 10^{5}~\mbox{TeV}
\end{equation}
this is indeed possible, as shown in the top panels of Fig.~\ref{fig:bulk}.
We recall that the bands correspond to the systematic uncertainty deriving from the ignorance of the running of each individual coupling of the gauge KK modes.

For smaller values of the compactification scale, it is not possible to determine the UV fate of the bulk Yukawa couplings. This is well illustrated by studying the RGE of the top Yukawa alone. Assuming that the other Yukawas remain small, the top RGE can be approximated in the UV as:
\begin{equation}
    2 \pi \frac{d \tilde{\alpha}_t}{dt} \approx \left[ 2 \pi + 15 \tilde{\alpha}_t - \frac{114}{5} \tilde{\alpha}_5 \right] \tilde{\alpha}_t\,.
\end{equation}
This as an UV fixed point for
\begin{equation}
    \tilde{\alpha}_t^\ast = \frac{1}{15} \left( \frac{114}{5} \tilde{\alpha}_5^\ast - 2 \pi\right) = \frac{(41 + 40 n_g) \pi}{75 (13-4 n_g)} = \frac{161}{75} \pi\,,
\end{equation}
where the numerical value is computed for 3 bulk generations.
By running down from this UV value to the EW scale, we predict a range for the top Yukawa which always includes the SM value. This implies that the top Yukawa may run to the fixed point, or to zero if the RGE trajectory lies below the critical one. This argument shows that the possibility that the theory is consistent for values of the compactification scale below the threshold in Eq.~\eqref{eq:limrun} is not excluded.

However, if the top Yukawa were to run to the UV fixed point, the other Yukawas would increase too fast at high energies. In fact, neglecting all Yukawas except the top and taking the top and gauge couplings at the UV fixed point, the high-energy RGEs can be approximated by
\begin{equation}
    2 \pi \frac{d \tilde{\alpha}_{b,\tau}}{dt} \approx \frac{399 \pi}{25} \tilde{\alpha}_{b,\tau}\,, \quad  2 \pi \frac{d \tilde{\alpha}_{\nu}}{dt} \approx \frac{469 \pi}{25} \tilde{\alpha}_{\nu}\,.
\end{equation}
Thus, the top Yukawa UV fixed point is excluded, and the only feasible possibility is that all bulk Yukawa couplings run to zero.
The possibility to salvage the interactive UV fixed point for the top, at the price of localizing some Yukawas or fermion fields, has been explored in Appendix~\ref{app:BulkTop}.

%%%%%%%%%%%%%%%%%%%%%%%%%%
\subsection{Numerical results for localized Yukawas}

Another possibility is that all Yukawa couplings are localized on a boundary of the extra dimension. For simplicity, and to preserve the GUT spirit, we will consider the case $y=\pi R/2$, where the $SU(5)$ symmetry is not broken. Thus, the localized Yukawa couplings have the same form of the bulk ones in Eq.~\eqref{Model-Lagrangian}. The linear running in this case is due to loops involving bulk fields, namely the gauge ones. By explicit calculation we find that the contribution is the same as for the bulk Yukawas in Eqs.~\eqref{eq:RGEtopbulk}--\eqref{eq:RGEneutrinobulk} except for a factor of two. Thus:
\begin{equation}
     2\pi \frac{d\alpha_f}{dt} = \left. 2\pi \frac{d\alpha_f}{dt} \right|_{\rm SM} +(S(t)-1)\ 2 \left[\begin{array}{c} - 114/5 \\ - 93/5  \\ - 93/5 \\ -6 \end{array} \right] \alpha_5\alpha_f \,,
\end{equation}
where $f = t, b, \tau, \nu$.

\begin{figure}[tb!]
\begin{centering}\begin{tabular}{cc}
 \phantom{xxxxxxxx} $\large \bf R^{-1} = 2.4~\mbox{TeV}$ & \phantom{xxxxxxxx} $\large \bf R^{-1} = 10^{10}~\mbox{TeV}$ \\    \includegraphics[scale=0.78]{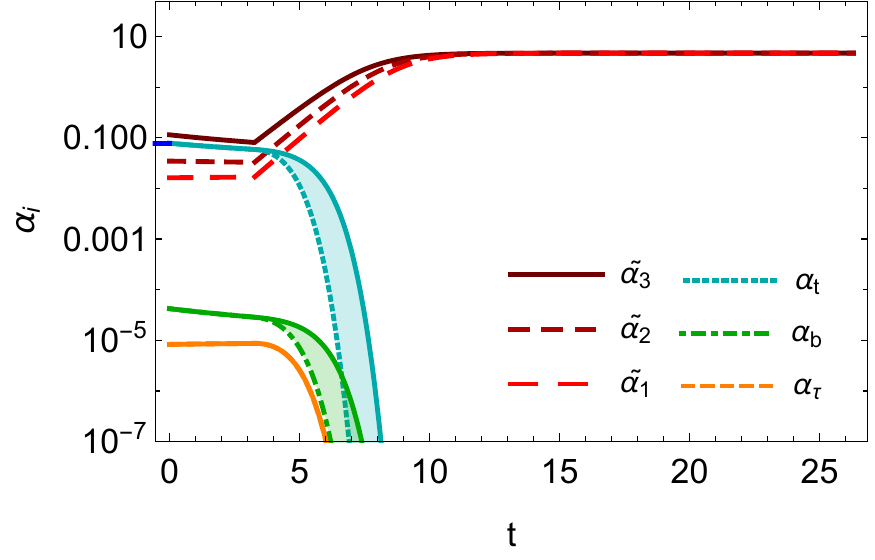} &
    \includegraphics[scale=0.78]{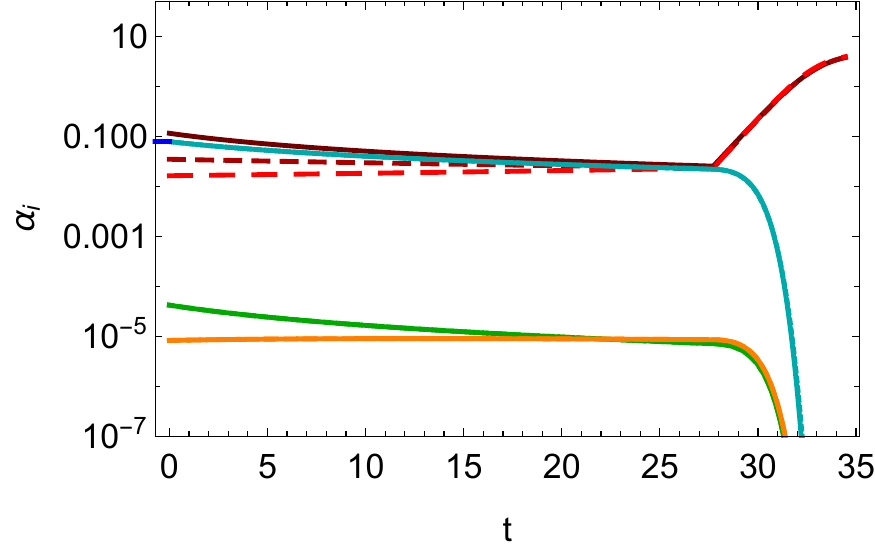} 
\end{tabular} \end{centering}
    \caption{Running of the localized Yukawa couplings compared to the bulk gauge ones for two sample values of the compactification scale. The bands indicate the uncertainty related to KK gauge couplings (see text). The largest value of $t$ corresponds to the 5D Planck mass value. }
    \label{fig:runloc}
\end{figure}

Numerically, we find that the localized Yukawas always run to zero in the UV, for any value of the compactification scale. Two examples of running are shown in Fig.~\ref{fig:runloc}, for two sample values of $R$. This configuration, therefore, always leads to a consistent $SU(5)$ aGUT model.

%%%%%%%%%%%%%%
\subsection{Implications for flavor physics, neutrino masses and a lower bound on $1/R$}

The structure of the Yukawa couplings discussed so far is the same as in the SM, in the sense of the flavor mixing. Thus, the scenarios where all Yukawas are either in the bulk or localized has the benefits of Minimal Flavor Violation, where strong flavor bounds are avoided \cite{Buras:2000dm,Bobeth:2005ck}. More specifically, for bulk Yukawa couplings, the flavor violation can only be generated by loops of the KK modes of the bulk fields. The suppression from the CKM matrix (being the sole source of flavor mixing) ensures that the bound on the compactification scale can be near the TeV scale. In the case of localized Yukawas, tree-level couplings also arise. The strongest bounds derive from $\epsilon_K$ in the Kaon sector and the mass mixing in the $B$ mesons, leading to a generic bound $\Lambda \gtrsim 6$~TeV~\cite{Isidori:2012ts}, which roughly translates on a bound on the KK mass. In Appendix~\ref{app:BulkTop} we also explore the possibility of having a bulk top Yukawa running to the UV fixed point, while other couplings are localized (in which case strong flavor bounds apply).

\par In the neutrino sector, masses are generated by the Yukawa coupling involving the singlet $\psi_1$. As Majorana masses are not allowed in 5 dimensions, bulk interactions can only generate Dirac masses for the SM neutrinos. On the other hand, a Majorana mass for the right-handed component of $\psi_1$ can be localized on either boundary. If this mass term is much larger than the compactification scale, a type-I see-saw can occur to suppress the Majorana masses of the left-handed neutrinos. Furthermore, this construction can generate leptogenesis \cite{Davidson:2008bu} at high scales, thus feeding the necessary asymmetry for the Indalogenesis and baryogenesis described in the next section. We leave a detailed study of this mechanism to future investigations.

%%%%%%%%%%%%%%%%%%%%%%%%%%%%%%%%%%%%%%%%%%%%
%  Section 5: Dark Matter phenomenology

\section{Indalo phenomenology and Dark Mattwer}\label{sec:5}

\par As discussed in the previous sections, the Indalo partners of the SM particles have non-standard baryon number assignments (see Table \ref{tab:2}), which prevent their decay into SM states alone. Thus, the lightest stable \Indalo -state is a natural candidate for Dark Matter. The mass splitting between the lowest tier states is induced by loops \cite{Cheng:2002iz} and by the effect of the Higgs. Thus, the lightest state is naturally the partner of the left-handed electron neutrino. Being part of a doublet, its annihilation and co-annihilation cross-sections suppress the thermal relic density, which is therefore insufficient to generate the necessary Dark Matter relic density. Furthermore, the interactions via the $Z$ boson are too strong to avoid exclusions by the null outcome of Direct Detection experiments \cite{Cui:2017nnn,Akerib:2016vxi,Aprile:2018dbl}. The left-handed neutrino \Indalo -partner, therefore, can only constitute a minor fraction of the total Dark Matter relic density in the present Universe.

\par The simplest way to  introduce a Dark Matter candidate is by doubling the singlet field, as done for the other GUT fermion fields. We thus introduce a second singlet, $\psi'_1$, with parities: 
\begin{eqnarray}\label{eqn:psi1P0}
    (P_0) &\Rightarrow& \left\{ 
    \psi'_1 (x,-y) = - \gamma_5 \psi'_1 (x,y) \,, \right. \\
\label{eqn:psi1P1}
    (P_1) &\Rightarrow& \left\{ 
    \psi'_1 (x,\pi R - y) = + \gamma_5 \psi'_1 (x, y) \, .
 \right.
\end{eqnarray}
Note that the second singlet allows us to combine the fields in each family into a $\bf 16$ and a ${\bf \bar{16}}$ of an $SO(10)$ symmetry that includes the gauged $SU(5)$. The parities allow us to add the new bulk Yukawa coupling:
\begin{equation}
\Delta \mathcal{L} = Y'_\nu \bar{\psi}'_1 \psi_{5} \phi_5^\ast + \mbox{h.c.}\, ,
\end{equation}
which could also be localized on the boundary. The new Yukawa obeys the following RGE:
\begin{equation}
    2\pi \frac{d\alpha_{\nu'}}{dt} =  (S(t)-1)
   \left[ 5 \alpha_{\nu'} + 12 \alpha_t + 18 \alpha_{b} + 8 \alpha_\tau + 2 \alpha_\nu - 6 \alpha_5 \right] \alpha_{\nu'} \,, \label{eq:RGEneutrinobulk2}
\end{equation}
while contributing $\{+2, +9/2, +2, +2\} \alpha_{\nu'}$ to the bulk RGEs of the top, bottom, tau and neutrino, respectively. For localized couplings, only the gauge contribution remains with a factor of 2. The modified RGEs still guarantee that the new Yukawa runs to zero in the UV.

The field $\psi'_1$ has no zero mode, and the lightest state corresponds to an \Indalo -singlet $\mathcal{S}$. The Yukawa coupling $Y'_\nu$ ensures that $\mathcal{S}$ has the same baryon and lepton numbers as the \Indalo -neutrino $\mathcal{N}$, see Table~\ref{tab:2}. 
Furthermore, the lightest \Indalo -state is now $\mathcal{S}$ because of the lack of gauge interactions that could lift its mass at loop level. Thus, any relic \Indalo -states will eventually decay into it via the Yukawa coupling $Y'_\nu$, which needs to be large enough to provide a lifetime shorter than the age of the Universe. Being a singlet, the Dark Matter candidate $\mathcal{S}$ does not suffer from strong Direct Detection constraints.

\par As all the \Indalo -states carry baryon and lepton number, a relic density can be generated at the EW phase transition together with the baryon asymmetry. This mechanism has been used to generate an asymmetric Dark Matter relic density \cite{Nussinov:1985xr}.

%%%%%%%%%%%%%%%%%%%%%%%%%%%%%%%%%%%%%%%%%%%%
\subsection{Indalogenesis via baryogenesis}\label{sec:5-1}

\par To estimate the relic density of \Indalo -particles produced during baryogenesis, we will rely on the usual calculation based on the equilibrium of the chemical potential for all species of particles active at the time of the EW phase transition \cite{Harvey:1990qw}. As we only want to provide an estimate, we will not analyze here the dynamics of the phase transition, which is complicated by the extra-dimensional nature of the theory \cite{Agashe:2020lfz}, and leave this study for future work. 
\begin{table}[tb!]
\begin{center}
\begin{tabular}{|ll|ll|ll|}
\hline
\multicolumn{2}{|c|}{$\phi_5$} & \multicolumn{2}{c|}{$\psi_{\overline{5}}$} & \multicolumn{2}{c|}{$\psi_{5}$} \\
\hline
$H$ :& $\mu_{H}$ & $B^c_R$ :& $- \mu_{B_R}$ & $b_R$ :& $\mu_{b_R}$ \\
$\phi^+$ :& $\mu_{+}$ & $\tau_L$ :& $\mu_{\tau_L}$ & $\mathcal{T}_L^c$ :& $- \mu_{\mathcal{T}_L}$ \\
$\phi_0$ :& $\mu_0$ & $\nu_L$ :& $\mu_{\nu_L}$ & $\mathcal{N}_L^c$ :& $- \mu_{\mathcal{N}_L}$ \\
\hline
\multicolumn{2}{|c|}{$\psi_{10}$} & \multicolumn{2}{c|}{$\psi_{\overline{10}}$} & \multicolumn{2}{c|}{$A^a_{\mu}$} \\
\hline
$T_R^c$ :& $- \mu_{T_R}$ & $t_R$ :& $\mu_{t_R}$ & $W^+$ :& $\mu_W$ \\
$\mathcal{T}^c_R$ :& $- \mu_{\mathcal{T}_R}$ & $\tau_R$ :& $\mu_{\tau_R}$ & $X$ :& $\mu_X$ \\
$t_L$ :& $\mu_{t_L}$ & $T_L^c$ :& $- \mu_{T_L}$ & $Y$ :& $\mu_Y$ \\
$b_L$ :& $\mu_{b_L}$ & $B_L^c$ :& $- \mu_{B_L}$ & & \\
\hline
\end{tabular}\end{center}
\caption{The chemical potentials associated to the relevant fields, where the SM ones correspond to the zero modes and the Indalo
ones to the lowest KK tier. \label{tab:6}}
\end{table}

\par The states we consider here, with their chemical potentials, are listed in Table~\ref{tab:6}, where the SM fields are associated to the zero modes and the \Indalo\ ones to the lowest tier of the KK modes. All other states are heavier and their contributions are neglected. Furthermore, we assume that the three families of fermions share the same chemical potentials. The gauge and Yukawa interactions impose many relationships (recapped in Appendix~\ref{app:IndaloGen}), which allow us to express all the chemical potentials in terms of 4 potentials, which we chose to be $\mu_{t_L}$, $\mu_{W}$, $\mu_{H}$ and $\mu_0$.

\par At the freeze-out temperature $T_{f}$, the matter-antimatter asymmetry for each species of mass $M$ can be written as:
\begin{equation}
    n = n_+ - n_- = d_{\mathrm{dof}}\, T_f^{3}\frac{\mu}{T_f}\frac{\sigma\left(\frac{M}{T_f}\right)}{6} \; ,
\end{equation}
where \cite{Gudnason:2006yj}
%\begin{widetext}
\begin{equation}
    \sigma(z) = \left\{
    \begin{array}{lll}
        \displaystyle \frac{6}{4\pi^2} \int^{\infty}_0 dx x^2 \text{cosh}^{-2}\left(\frac{1}{2}\sqrt{x^2 + z^2}\right) & \mbox{for fermions,}  \\
        \\
       \displaystyle \frac{6}{4\pi^2} \int^{\infty}_0 dx x^2 \text{sinh}^{-2}\left(\frac{1}{2}\sqrt{x^2 + z^2}\right) & \mbox{for bosons.} 
    \end{array}
\right.
\end{equation}
%\end{widetext}
The $\sigma$ function is normalized to 1 for massless fermions and to 2 for massless bosons. In the following we will consider that only the \Indalo -states and the top quark have a non-negligible mass, so that the particle density will be set equal to 1 for fermions and 2 for bosons. We have summarized the total density of each species in Table~\ref{tab:7}. Moreover, the charges and the iso-spins of all relevant states can be recapped in Table~\ref{tab:2}.
\begin{table}[tb]
\begin{center}
\begin{tabular}{ |p{.8cm}|p{3.5cm}||p{.8cm}|p{3cm}| }
 \hline
 Field    & Density  & Field    & Density\\
 \hline
 $t$ & $3(2+\sigma_t)(\mu_{t_L}+\mu_{t_R})$ & $b$ & $9(\mu_{b_L}+\mu_{b_R})$\\
  \hline
 $\nu$ & $3(\mu_{\nu_L})$ & $\phi^-$ & $2\mu_{\phi^-}$\\
   \hline
  $\tau$ & $3(\mu_{\tau_L}+\mu_{\tau_R})$ & $h$ & $2\mu_h$\\
 \hline
  $T$ & $18 \sigma_{T} (\mu_{T_L}+\mu_{T_R})$ &  $B$ & $18\sigma_{B} (\mu_{B_L}+\mu_{B_R})$\\
  \hline
  $\mathcal{N}$ & $6 \sigma_{\mathcal{N}}\mu_{\mathcal{N}_L}$  & $\mathcal{T}$ & $6 \sigma_{\mathcal{T}} (\mu_{\mathcal{T}_L}+\mu_{\mathcal{T}_R})$ \\
 \hline
 $X$  & $3\sigma_X\mu_X$ & $Y$ & $3\sigma_Y\mu_Y$ \\
 \hline
  $H$ & $3\sigma_H\mu_H$ & & \\
  \hline
\end{tabular} \end{center}
\caption{The normalized particle densities.}
\label{tab:7}
\end{table}

\par With these ingredients we can calculate, in a straightforward way, the total baryon number stored separately in the SM and in the \Indalo -sectors. At any given temperature they will depend on the total densities of each species, cf. Table~\ref{tab:7}. After the \Indalo -particles exit thermal equilibrium with the SM, they will promptly decay into the lightest one, $\mathcal{S}$, and release some baryon number to the SM sector again. For instance, we see in Table~\ref{tab:2} that the \Indalo -quarks $T$ and $B$ have baryon number $-1/6$, while $\mathcal{S}$ has baryon number $-1/2$, thus the decay will release a net baryon number $1/3$ to the SM sector. The same consideration applies to the bosons $H$, $X$ and $Y$, while $\mathcal{T}$ and $\mathcal{N}$ have the same baryon number as $\mathcal{S}$.  Finally, after freeze-out, the baryon numbers in SM baryons and in $\mathcal{S}$ can be expressed as:
\begin{eqnarray}
B_{\rm SM} &=& \frac{1}{3}3(2+\sigma_t)(\mu_{t_L} + \mu_{t_R}) + \frac{1}{3}9(\mu_{b_L} + \mu_{b_R}) 
+  \frac{1}{3}3(\sigma_H\mu_H + \sigma_X\mu_X + \sigma_Y\mu_Y)  \nonumber \\ 
&+&  \frac{1}{3} 18 \sigma_B (\mu_{B_L} + \mu_{B_R}) + \frac{1}{3} 18 \sigma_T (\mu_{T_L} + \mu_{T_R})\,,
\end{eqnarray}
\begin{eqnarray}
B_{\mathcal{S}} &=& -\frac{1}{2}( 3\sigma_{\mathcal{N}}(\mu_{\mathcal{N}_L}) + 3\sigma_{\mathcal{T}}(\mu_{\mathcal{T}_L}+\mu_{\mathcal{T}_R}) 
+  3(\sigma_H\mu_H + \sigma_X\mu_X + \sigma_Y\mu_Y) \nonumber \\
&+&   9 \sigma_B (\mu_{B_L} + \mu_{B_R}) + 9 \sigma_T (\mu_{T_L} + \mu_{T_R}))\,.
\end{eqnarray}
Using the relationships among the chemical potentials, and assuming for simplicity that all \Indalo -states have the same mass, i.e. $\sigma_T = \sigma_B = \sigma_{\mathcal{T}} = \sigma_{\mathcal{N}} \equiv \sigma_F$ and $\sigma_X = \sigma_Y = \sigma_H$, we find:
\begin{equation}
    B_{\rm SM} =  (10 + 2\sigma_t - 24\sigma_F)\mu_{t_L} + (12\sigma_F - 6 - \sigma_X)\mu_W +  (3\sigma_X - 24\sigma_F)\mu_H + (\sigma_t - 1)\mu_0\,,
\end{equation}
\begin{equation}
B_{\mathcal{S}} = \frac{3}{2}\left(30\sigma_F\mu_{t_L}  +(\sigma_X - 14\sigma_F)\mu_W  +  (18\sigma_F - 3\sigma_X)\mu_H + 2\sigma_F\mu_0 \right)\,.
\end{equation}
The mass density of $\mathcal{S}$ divided by the baryon density can now be expressed as
\begin{equation}
\frac{\Omega_{\mathcal{S}}}{\Omega_b} = \frac{2 m_\mathcal{S}\ B_\mathcal{S}}{m_p\ B_{\rm SM}}\,,
\end{equation}
where $m_p$ is the proton mass.

\par The dependence on the chemical potentials can be further reduced by considering the EW phase transition, where we assume that it can be of 1st or 2nd order. As all \Indalo -states are vector-like, they do not contribute to the sphaleron rate, so that the same relation holds as in the SM \cite{Harvey:1990qw}:
\begin{equation}
    3 (\mu_{t_L} + 2\mu_{b_L}) + 3\mu_{\nu_L} =  3\mu_{t_L} - 2\mu_W + 2\mu_H = 0\,.
\end{equation}
The other relevant quantities are the total electric charge and the total weak iso-spin (see Appendix \ref{app:IndaloGen}). Under the same approximations as above, they read:
\begin{eqnarray}
    Q^{tot} &=&  4(\sigma_t-1)\mu_{t_L} + (10 + 24\sigma_F + 4\sigma_X)\mu_W  - (12 + 2\sigma_F + 4\sigma_X)\mu_H   \nonumber \\
     &+& (12 + 2\sigma_t + 24\sigma_F - 3\sigma_X)\mu_0\,, \\
    Q_3^{tot} &=& 3(\sigma_t - 1)\mu_{t_L} + (10 + 12\sigma_F + 3\sigma_X)\mu_W - 6\sigma_X\mu_0\,,
\end{eqnarray}
respectively.

%%%%%%%%%%%%%%%%%%%%%%%%%%%%%%%%%%%%%%%%%%%%
\subsection{Numerical results}\label{sec:5-2}

\par A first order phase transition is characterized by the vanishing of the total charge and weak iso-spin, $Q^{tot} = Q_3^{tot} = 0$. Together with the sphaleron condition, they allow us to write the baryon numbers in terms of a single chemical potential. Analogous results can be obtained for a second order phase transition, where the vanishing of the weak iso-spin condition is replaced by the vanishing of the Higgs chemical potential, i.e. $\mu_0 = 0$.

\begin{figure}[ht!]
\begin{centering}\begin{tabular}{cc}
{\large \bf \phantom{xxxx} 1${}^{\rm st}$ order phase transition} & {\large \bf \phantom{xxxx} 2${}^{\rm nd}$ order phase transition} \\
\includegraphics[scale=0.57]{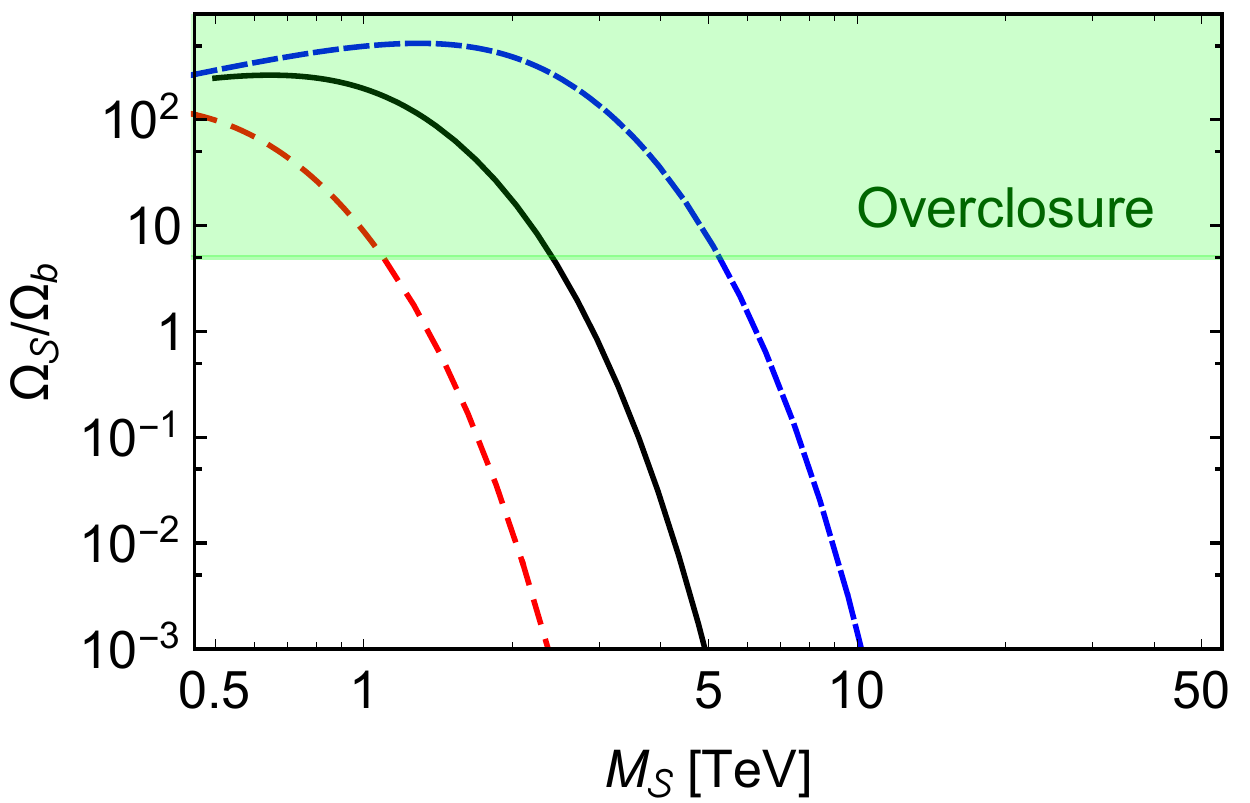} & \includegraphics[scale=0.57]{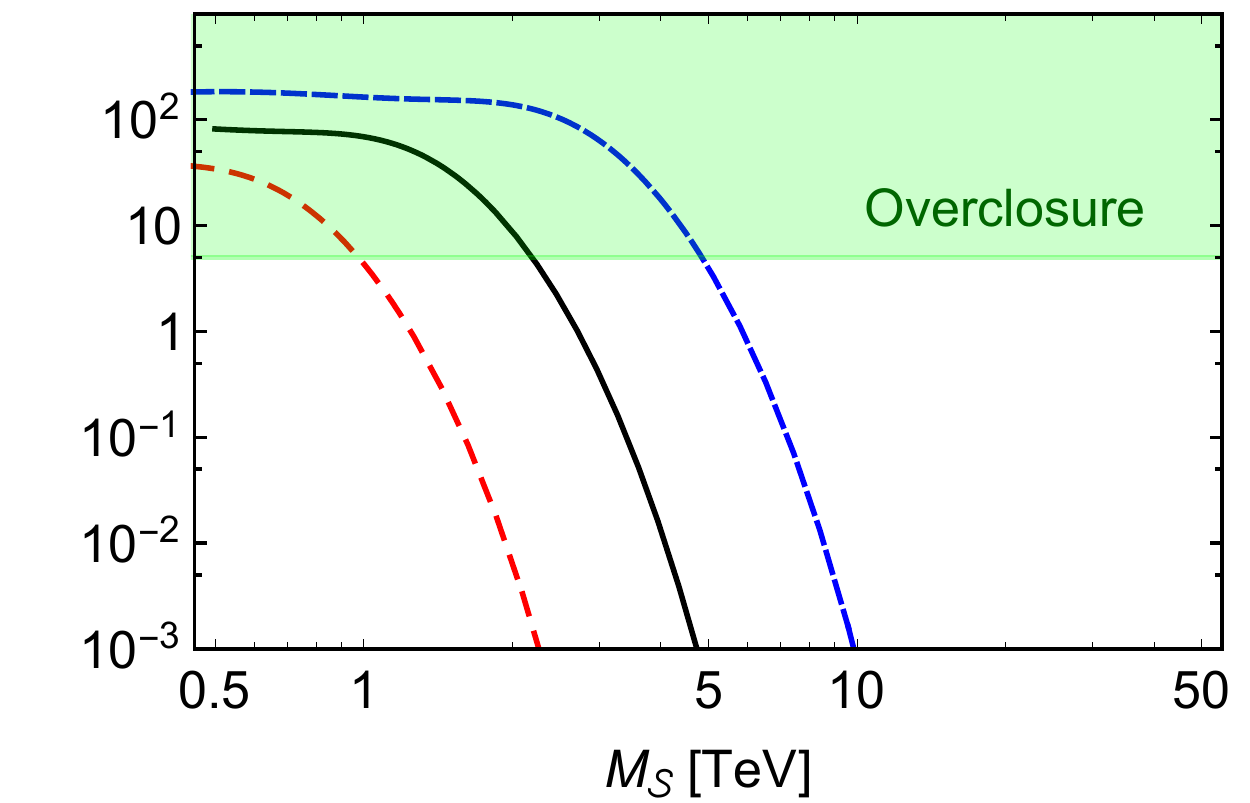}
\end{tabular}\end{centering}
    \caption{1$^{\rm st}$ order (left panel) and 2$^{\rm nd}$ order (right panel) phase transition results. Values of the $\mathcal{S}$ relic density as a function of the mass for $T^\ast = v_{\rm SM}$ (solid black), $v_{\rm SM}/2$ (dashed red) and $2 v_{\rm SM}$ (dashed blue). The green shaded region is excluded by the overclosure of the Universe.}
    \label{fig:4}
\end{figure}

\par The ratio of relic densities, therefore, only depends on the $\mathcal{S}$ mass and the temperature of the phase transition. The solutions are shown in Fig.~\ref{fig:4}, where we plot the relic density as a function of the mass for three values of the EW phase transition temperature. The region in green is excluded by the overclosure of the Universe, thus this mechanism imposes a lower bound on the KK mass scale. For the Indalo $\mathcal{S}$ state to saturate the relic density, the mass should lie within the range 1 to 6~TeV depending on the temperature of the EW phase transition, where the central value is
\begin{equation}
    m_{KK} \approx 2.4~\mbox{TeV for}\;\; T^\ast = v_{\rm SM}\,.
\end{equation}

%%%%%%%%%%%%%%%%%%%%%%%%%%%%%%%%%%%%%%%%%%%%
%  Section 7: Conclusions

\section{Conclusions}\label{sec:7}

\par We propose a concrete 5-dimensional model that realizes asymptotic Grand Unification for the gauge couplings. The model is based on a bulk $SU(5)$ gauge symmetry, broken on a $S^1/(\mathbb{Z}_2 \times \mathbb{Z}'_2)$ orbifold. Due to the orbifold parity assignments, the SM fermions cannot be embedded as zero modes of a single family of $\bf 10 \plus \bar{5}$. As a consequence, the Yukawa couplings do not unify at high energies. We find that, in the minimal aGUT model, asymptotic unification is only possible for a number of bulk generations less than or equal to 3. The effective 5D gauge couplings thus run to a fixed point in the ultra-violet.

\par We also studied the running of the Yukawa couplings, assuming that they are all in the bulk or localized on the $SU(5)$ boundary. We showed that they are guaranteed to run asymptotically free in the localized case and for large compactification scales in the bulk case ($1/R \gtrsim 3 \times 10^5$~TeV). For smaller scales,  the possibility of an asymptotically free top Yukawa cannot be excluded. Flavor mixing can be introduced by promoting the Yukawa couplings to matrices, within a natural Minimal Flavor Violation scenario that reduces the flavor bounds on the compactification scale to the TeV scale.

\par The non-unification of the SM Yukawas implies that baryon and lepton numbers are preserved. Furthermore, we find that all the field components without a zero mode have $B$ and $L$ values which are half of those of the SM particles.  The implication is two-fold: proton decay is avoided, thus the compactification scale and unification can occur at low energies; the lightest KK state is stable and a potential Dark Matter candidate is present. For this reason we suggested for them the name Indalo (\Indalo). A realistic scenario is achieved in the presence of an \Indalo -singlet, corresponding to the right-handed neutrinos, whose relic density is produced via Indalogenesis and baryogenesis at the EW phase transition. We estimate the Dark Matter mass, equal to the inverse radius, to lie in the range 1 to 6~TeV, depending on the temperature of the phase transition, with a value of $2.4$~TeV for the Higgs vacuum expectation value. Collider bounds apply to the second tier states, which have a mass that is double when compared to the \Indalo -states. As they can be singly-produced only at one loop, the bounds from the LHC are too weak. They could be accessible at future hadron colliders, especially in the mass range relevant for Indalo Dark Matter.

\par In summary, we presented a minimal aGUT model in 5 dimensions, which incarnates a new asymptotic unification paradigm. The presence of UV fixed points for the bulk gauge and top Yukawa couplings requires \emph{maximally} 3 bulk generations. The physics of neutrinos is also rich, allowing for Indalogenesis and a Dark Matter candidate. 

%%%%%%%%%%%%%%%%%%%%%%%%%%%%%%%%%%%%%%%%%%%%
%  Acknowledgements

\section*{Acknowledgements}
 ASC is supported in part by the National Research Foundation of South Africa (NRF) and thanks the University of Lyon 1 and IP2I for support during his collaborative visit to Lyon.
 
\appendix

\section{Models with top bulk Yukawa running to an UV fixed point}
\label{app:BulkTop}

We have seen that, neglecting the effect of the other Yukawas, the bulk top RGE has an interactive UV fixed point, given by: 
\begin{equation}
\tilde\alpha_t^\ast = \frac{114}{75}\tilde\alpha_5^\ast - \frac{2\pi}{15} = \frac{41+40 n_g}{75 (13-4 n_g)} \pi = \left\{ \begin{array}{cc}
\frac{3 \pi}{25}\,, \phantom{\Big(} & n_g=1 \\
\frac{121 \pi}{375}\,, \phantom{\Big(}  & n_g=2 \\\frac{161 \pi}{75}\,, \phantom{\Big(}  & n_g=3
\end{array} \right.\,.
\end{equation}
It is tantalizing to explore the model building route where this feature is preserved, while other Yukawas are allowed to be either localized or in the bulk.

%%%%%%%%%%%%%%%%%%%%%%%
\subsection{Bulk bottom and tau Yukawa couplings}

We first assume that only the neutrino Yukawas (both $Y_\nu$ and $Y'_\nu$) are localized.
To study the UV fate of the bottom and tau Yukawa couplings in the bulk, we can first analyze the UV asymptotic behavior of their beta functions under the assumption that the tau and bottom Yukawa themselves remain small. The RGEs in Eqs.~\eqref{eq:RGEbottombulk} and \eqref{eq:RGEtaubulk}, once the gauge and top Yukawa couplings have reached their fixed point values, can be approximated as
\begin{equation}
2 \pi \frac{d\tilde\alpha_{b,\tau}}{dt} = \left(2 \pi + \frac{39}{2} \tilde\alpha_t^\ast - \frac{93}{5} \tilde\alpha_5^\ast \right) \tilde\alpha_{b,\tau} = \frac{3 (73 + 20 n_g) \pi}{75 (13-4 n_g)} \tilde\alpha_{b,\tau}\,,
\end{equation}
where we neglected higher orders in the bottom and tau Yukawa couplings. The positivity of the coefficient implies that the two couplings will grow as a power-law of the energy in the UV, thus they will not remain small and negligible for long. In reverse, starting from small couplings at the 5D Planck mass would imply unrealistically suppressed values at the EW scale.

It is, therefore, necessary to study the eventual presence of fixed points for the three coupled RGEs in Eqs~\eqref{eq:RGEtopbulk}--\eqref{eq:RGEtaubulk}, taken in the asymptotic limit. We find
\begin{equation}
\alpha_t^\ast = \frac{14 (10 n_g - 19)}{555 (13-4 n_g)} \pi\,, \qquad \alpha_b^\ast = \alpha_\tau^\ast = \frac{3 (73 + 20 n_g) \pi}{370 (13-4 n_g)}\,.
\end{equation}
Interestingly, the top fixed point remains positive only for 2 and 3 bulk generations, while it turns negative for $n_g = 1$, thus signalling an instability. However, the fixed point for the top coupling is always smaller than that for the bottom and tau, $\alpha_t^\ast < \alpha_b^\ast = \alpha_\tau^\ast$.  
Running down from the fixed point, therefore, the top Yukawa undershoots the SM value at the EW scale, while the other two remain unrealistically large. We have checked this point numerically for both 3 and 2 bulk generations. 

This simple asymptotic analysis shows that a model with top, bottom and tau Yukawa couplings in the bulk is unrealistic, when the top flows to its interactive UV fixed point, as it cannot reproduce the SM at low energies.

%%%%%%%%%%%%%%%%%%%%%%%
\subsection{Localized $SU(5)$-invariant bottom and tau Yukawa couplings}

\par For simplicity, and to preserve the GUT spirit, we will first consider the case $y=\frac{\pi R}{2}$, where the $SU(5)$ symmetry is not broken. Thus, the localized Yukawa couplings have the same form of the bulk ones in Eq.~\eqref{Model-Lagrangian}. The linear running is due to loops involving bulk fields, namely the gauge and top Yukawa, leading to:
\begin{eqnarray}
     2\pi \frac{d\alpha_{b,\tau}}{dt} &=& \left. 2\pi \frac{d\alpha_{b,\tau}}{dt} \right|_{\rm SM} +(S(t)-1)\ 2 \left[ \frac{39}{2} \alpha_t - \frac{93}{5} \alpha_5 \right] \alpha_{b,\tau} \,, \\
     2\pi \frac{d\alpha_{\nu}}{dt} &=& \left. 2\pi \frac{d\alpha_\nu}{dt} \right|_{\rm SM} +(S(t)-1)\ 2 \left[ 12 \alpha_t - 6 \alpha_5 \right] \alpha_\nu \,, \\
    2\pi \frac{d\alpha_{\nu'}}{dt} &=& (S(t)-1)\ 2 \left[ 12 \alpha_t - 6 \alpha_5 \right] \alpha_{\nu'} \,.
\end{eqnarray}

\par We first study the UV behavior of the RGEs under the assumption that the localized couplings remain small and negligible. For gauge and top Yukawas at the bulk fixed point, we find:
\begin{equation}
    2 \pi \frac{d \alpha_{b,\tau}}{dt} = - \frac{(862 - 520 n_g) \pi}{25 (13-4 n_g))} \alpha_{b,\tau}\,, \quad 2 \pi \frac{d \alpha_{\nu,\nu'}}{dt} = \frac{(160 n_g - 61) \pi}{25 (13-4 n_g))} \alpha_{\nu,\nu'}\,.
\end{equation}
Interestingly, for $n_g = 1$, i.e. one bulk generation, the bottom and tau RGEs are negative and the corresponding Yukawas run free in the UV. On the other hand, the RGEs for the neutrinos are always positive, signalling that the respective Yukawas are suppressed at low energies. For $n_g = 1$:
\begin{equation}
    y_{\nu,\nu'} (1/R) \approx y_{\nu,\nu'} (M_{\rm Pl}^\ast)\times \left( \frac{1}{M_{\rm Pl}^\ast R}\right)^{11/50}\,.
\end{equation}
Due to the relatively small exponent, sizeable values of the neutrino Yukawas at the compactification scale can be achieved, starting from $\mathcal{O} (1)$ values at the reduced Planck scale.

This is, therefore, a realistic scenario only for 1 bulk generations. The main drawback of this scenario is due to large tree-level flavor violating effects, which arise due to the different nature of the top quark Yukawa versus the other quarks. This implies that the compactification scale needs to be very large, and Cosmological production of the Indalo states need to be suppressed in order to avoid overclosure of the Universe. Another issue is that to implement the Yukawa couplings of the light two generations, which are constituted by localized fields, one would need to introduce corresponding chiral and massless $SU(5)$ partner states, which are inconsistent with the SM.

%%%%%%%%%%%%%%%%%%%%%%%
\subsection{Bottom and tau Yukawa couplings localized on the $SU(5)$-breaking boundary}

To avoid the issue with the localized $SU(5)$ multiplets, we explore here the possibility that all Yukawas but the top one are localized on the $y=0$ boundary, where $SU(5)$ is broken to the SM gauge group. The Yukawas, therefore, have the same structure as the SM ones and no additional chiral fields are needed except the SM ones for the two light generations.
The RGEs read:
\begin{eqnarray}
     2\pi \frac{d\alpha_b}{dt} &=& \left. 2\pi \frac{d\alpha_b}{dt} \right|_{\rm SM} +(S(t)-1)\ 2 \left[ \frac{3}{2} \alpha_t - \frac{1}{4} \alpha_1 - \frac{9}{4} \alpha_2 - 8 \alpha_3 \right] \alpha_b \,, \\
     2\pi \frac{d\alpha_\tau}{dt} &=& \left. 2\pi \frac{d\alpha_\tau}{dt} \right|_{\rm SM} +(S(t)-1)\ 2 \left[ 3 \alpha_t - \frac{9}{4} \alpha_1 - \frac{9}{4} \alpha_2 \right] \alpha_\tau \,, \\
     2\pi \frac{d\alpha_\nu}{dt} &=& \left. 2\pi \frac{d\alpha_\nu}{dt} \right|_{\rm SM} +(S(t)-1)\ 2 \left[ 3 \alpha_t - \frac{9}{20} \alpha_1 - \frac{9}{4} \alpha_2 \right] \alpha_\tau \,.
\end{eqnarray}
As before, we can compute the asymptotic RGEs at leading order in the small Yukawas, leading to
\begin{eqnarray}
    & 2 \pi \frac{d\alpha_b}{dt} = - \frac{(1493-80 n_g)\pi}{50 (13-4 n_g)} \alpha_b\,, \quad     2 \pi \frac{d\alpha_\tau}{dt} = - \frac{(511-160 n_g)\pi}{50 (13-4 n_g)} \alpha_\tau\,, & \\
    & 2 \pi \frac{d\alpha_\nu}{dt} = - \frac{(241-160 n_g)\pi}{50 (13-4 n_g)} \alpha_\nu\,. &
\end{eqnarray}
The RGEs for the bottom and tau are negative for any number of bulk generations up to 3, while the neutrino one turns positive for 2 and 3 bulk generations. This implies that the bottom and tau Yukawas run to a free asymptotic value in the UV. 
For the neutrino Yukawa, a suppression in the IR arises for 2 and 3 bulk generations, however still leading to reasonable values at the compactification scales:
\begin{equation}
    y_{\nu,\nu'} (1/R) \approx y_{\nu,\nu'} (M_{\rm Pl}^\ast)\times \left( \frac{1}{M_{\rm Pl}^\ast R}\right)^{x}\,, \qquad x = \left\{ \begin{array}{l}
    79/1000\;\; \mbox{for}\;\; n_g=2 \\
     239/200\;\; \mbox{for}\;\; n_g=3
    \end{array}\right.
\end{equation}
This scenario, therefore, allows for a model with 3 bulk generations. However, flavor violation due to the different nature of the top Yukawa from the others would still push the value of the compactification scale to large scales.

\section{Details of the Indalogenesis}
\label{app:IndaloGen}

\par According to the interactions of the Lagrangian in Eq.~\eqref{Model-Lagrangian}, the following relationships among the chemical potentials hold:
\begin{eqnarray}
\mu_{H} &=& - \mu_{T_R} - \mu_{b_R} =  \mu_{b_L} + \mu_{\mathcal{N}_L} =  \mu_{t_L} + \mu_{\mathcal{T}_L}   =   - \mu_{B_R} - \mu_{t_R} = \mu_{\tau_L} + \mu_{T_L} \notag \\
& = &  \mu_{\nu_L} + \mu_{B_L} = \mu_{\tau_R} + \mu_{T_R} = \mu_{t_R} + \mu_{\mathcal{T}_R} =  - \mu_{T_L} - \mu_{b_L} = - \mu_{B_L} - \mu_{t_L} \; ,
\end{eqnarray}
\begin{equation}
\mu_{0}  =  \mu_{b_L} - \mu_{b_R} = \mu_{\mathcal{T}_L} - \mu_{\mathcal{T}_R} =  \mu_{T_R} - \mu_{T_L} 
= \mu_{t_R} - \mu_{t_L} = \mu_{B_L} - \mu_{B_R} =  \mu_{\tau_L} - \mu_{\tau_R} \; , 
\end{equation}
\begin{equation}
\mu_{+}  =  \mu_{\mathcal{N}_L} - \mu_{\mathcal{T}_R} = \mu_{t_R} - \mu_{b_L} = \mu_{t_L} - \mu_{b_R} 
 =  \mu_{T_L} - \mu_{B_R} = \mu_{\nu_L} - \mu_{\tau_R} \; , 
\end{equation}
from the Yukawa couplings, and:
\begin{equation}
\mu_W = \mu_{T_L} - \mu_{B_L} = \mu_{\nu_L} - \mu_{\tau_L} = \mu_{\mathcal{N}_L} - \mu_{\mathcal{T}_L}
= \mu_{t_L} - \mu_{b_L}  =  \mu_{+} - \mu_{0} \; , 
\end{equation}
\begin{equation}
\mu_X = \mu_{\tau_L} + \mu_{B_R} = \mu_{b_R} + \mu_{\mathcal{T}_L} = -\mu_{t_L} - \mu_{T_R} 
= \mu_{b_L} + \mu_{\mathcal{T}_R} = -\mu_{t_R} - \mu_{T_L} = \mu_{\tau_R} + \mu_{B_L} \; ,
\end{equation}
\begin{equation}
\mu_Y =  \mu_{\nu_L} + \mu_{B_R} = \mu_{b_R} + \mu_{\mathcal{N}_L} = -\mu_{b_L} - \mu_{\mathcal{T}_R}
  =  \mu_{t_L} + \mu_{\mathcal{T}_R} = -\mu_{t_R} - \mu_{B_L} = \mu_{\tau_R} + \mu_{T_L}\,, 
\end{equation}
from the gauge vertices.

\par The total charge density of the Universe and the weak iso-spin density are given by:
\begin{eqnarray}
Q^{Tot} &=& \frac{2}{3}3(2+\sigma_t)(\mu_{t_L} + \mu_{t_R}) - \frac{1}{3}(3)(3)(\mu_{b_L} + \mu_{b_R}) -  \sum_i(\mu_{\tau^i_L} + \mu_{\tau^i_R}) - 2 (2) \mu_W + 2\mu_{+} \notag \\ 
&& - 1 (3 \sigma_{\mathcal{T}}) (\mu_{\mathcal{T}_L} + \mu_{\mathcal{T}_R}) - \frac{1}{3} (3) (3 \sigma_B) (\mu_{B_L} + \mu_{B_R}) \notag \\
&& + \frac{2}{3} (3) (3 \sigma_T) (\mu_{T_L} + \mu_{T_R}) - \frac{4}{3} (3)\sigma_X \mu_X - \frac{1}{3} (3) \sigma_Y \mu_Y - \frac{1}{3}(3) \sigma_H \mu_H\,,
\end{eqnarray}
\begin{equation}
\begin{aligned}
Q_3^{Tot} =& 3(2+\sigma_t)(\mu_{t_L}) - 9(\mu_{b_L}) - \sum_i(\mu_{\tau^i_L} - \mu_{\nu^i_L}) - 4 \mu_W + 2(\mu_{+}-\mu_{0}) 
+ 6(\mu_{\mathcal{T}_L}\sigma_{\mathcal{T}}-\mu_{\mathcal{N}_L}\sigma_{\mathcal{N}}) 
\\ & + 18(\mu_{T_L}\sigma_{T}-\mu_{B_L}\sigma_{B}) + 3 (\sigma_Y\mu_Y - \sigma_X\mu_X) \,.
\end{aligned}
\end{equation}

%%%%%%%%%%%%%%%%%%%%%%%%%%%%%%%%%%%%%%%%%%%%
%  References

\bibliographystyle{JHEP-2-2}
\bibliography{biblio.bib}

\end{document}